\documentclass[12pt]{amsart}
\usepackage{amsmath,amssymb,amsthm}

\newtheorem{lemma}{Lemma}
\newtheorem{proposition}{Proposition}
\newtheorem{theorem}{Theorem}

\theoremstyle{definition}
\newtheorem{remark}{Remark}
\newtheorem{example}{Example}
\newtheorem{definition}{Definition}

\newcommand{\kb}[2]{\left|#1\right\rangle\left\langle#2\right|}
\newcommand{\ket}[1]{\left|#1\right\rangle}

\newcommand{\Z}{\mathbb Z}

\newcommand{\N}{\mathbb N}
\newcommand{\C}{\mathbb C}
\newcommand{\R}{\mathbb R}

\newcommand{\torus}{\mathbb T}

\newcommand{\lto}{L_2(\torus,\mu)}
\newcommand{\ltok}{L_2(\torus\times\torus,\mu\times\mu)}

\newcommand{\fii}{\varphi}
\newcommand{\hi}{\mathcal H}
\newcommand{\lh}{\mathcal{L(H)}}
\newcommand{\lhh}{\mathcal{L(H\otimes H)}}

\newcommand{\bor}[1]{\mathcal{B}({#1})}
\newcommand{\bo}{\mathcal B([0,2\pi))}
\newcommand{\boto}{\mathcal B(\torus)}

\newcommand{\Ediff}{E^{\rm diff}}
\newcommand{\Ecan}{\Ediff_{\rm can}}
\newcommand{\Ec}{E_{\rm can}}

\newcommand{\ud}{\textrm{d}}

\newcommand{\st}[2]{\langle {#1}|{#2}\rangle} 
\newcommand{\ketbra}[2]{|{#1}\rangle\langle {#2}|}
\newcommand{\ketbrab}[2]{|{#1}\rangle\rangle\langle\langle {#2}|}

\begin{document}

\sloppy

\title[Covariant phase difference observables]{Covariant phase difference observables}
\author{T.\ Heinonen}
\address{Teiko Heinonen, Department of Mathematics, University of Turku,
20014 Turku, Finland}
\email{temihe@utu.fi}
\author{P.\ Lahti}
\address{Pekka Lahti, Department of Physics, University of Turku,
20014 Turku, Finland}
\email{pekka.lahti@utu.fi}
\author{J.-P.\ Pellonp\"a\"a}
\address{Juha-Pekka Pellonp\"a\"a, Department of Physics, University of Turku,
20014 Turku, Finland}
\email{juhpello@utu.fi}
\date{\today}
\begin{abstract}
Covariant phase difference observables are determined in two different ways,
by a direct computation and by a group theoretical method.  A characterization of phase difference observables which can be expressed as difference of two phase observables is given.
Classical limit of such phase difference observables are determined and the Pegg-Barnett phase difference
distribution is obtained from the phase difference representation. The relation of Ban's theory to the covariant phase theories is exhibited.
\end{abstract}
\maketitle

\section{Introduction}
In 
quantum optical phase measurements
like heterodyne and eight-port homodyne detections one can measure
the phase difference between
two single-mode input fields.
However, if the second field, the reference field, can
be considered as a classical field with  well-known phase and (high) amplitude,
then the theory reduces to a single-mode theory with one input beam.
Under such conditions
 the heterodyne \cite{ShWa,Ha2} 
and the  eight-port homodyne
\cite{FrVoSc,LePa1,BuGrLa1} detection schemes 
measure the single-mode phase observable
$$
X\mapsto E_{|0\rangle}(X):=\frac{1}{\pi}\int_X\int_0^\infty|z\rangle\langle z|\,
|z|\,\mathrm d|z|\,\mathrm d{(\arg z)}
$$
defined in terms of the  coherent states
$|z\rangle:=e^{-|z|^2/2}\sum_{n=0}^\infty
z^n/\sqrt{n!}\,|n\rangle$. Here  $|z\rangle\langle z|$ denotes the
projection on the one-dimensional subspace spanned by $|z\rangle$
and $z= |z|\,\arg z$ is a complex number. The phase observable
$E_{|0\rangle}$ is covariant with respect to the shifts generated
by the single-mode number operator $N:=\sum_{n=0}^\infty n
|n\rangle\langle n|$, that is,
$$
e^{i\theta N}E_{|0\rangle}(X)e^{-i\theta N}=E_{|0\rangle}(X\dotplus\theta)
$$
for all (Borel) sets $X\subseteq[0,2\pi)$ and $\theta\in\R$, with
$\dotplus$ denoting  the addition modulo $2\pi$.
This condition is a natural covariance condition for
observables describing coherent state phase measurements,
and one may define a (single-mode) phase observable as a
phase shift covariant normalized positive operator measure
\cite{Holevo,OQP,LaPe2,PBCov}.
The structure of such observables is completely known and they can be
characterized in at least four different ways
in terms of
phase matrices, sequences of unit vectors, sequences of generalized vectors,
or using covariant trace-preserving operations, see e.g.\ \cite{Holevo,LaPe2,CDLP,Pell2}.

In this paper we consider the difference  of
two (single-mode)
phase observables and we notice that it
satisfies a natural covariance condition. We take this condition as the
defining condition of (two-mode) phase difference observables.
We give both a direct (Sect. \ref{direct}) and a group theoretical (Sect. \ref{covdis})
characterization of such observables whereas in Section \ref{diffe} we obtain a characterization of
the phase difference observables which can be expressed as a difference of two phase observables.
Section 6 puts the phase difference distribution of Barnett and Pegg \cite{BP90,PB97} in the present context.
 Section 7 studies the classical limit of  the two-mode theory  whereas Section 8 discusses the relation of Ban's theory \cite{Ban1,Ban2,Ban3} to the covariant phase and phase difference theories. In the final section of the paper some historical remarks are due and the question of
measurability of the phase difference is briefly reviewed. 

\section{Phase difference observables}
Two phase observables can be joined in a natural way to a phase difference observable.
Indeed, suppose that $E_1:\bor{[0,2\pi)} \to \lh$ and $E_2:\bor{[0,2\pi)} \to \lh$ are phase observables,
where $\bor{[0,2\pi)}$ is the Borel $\sigma$-algebra  of the phase interval
$[0,2\pi)$,  $\hi$ is a separable Hilbert space spanned by the number states $|n\rangle$,
$n\in\N$, and $\mathcal{L}(\hi)$ is the set of bounded operators on $\hi$.
The product map $(X,Y)\mapsto E_1(X)\otimes E_2(Y)$ defines a unique
 operator measure
$$
\tilde{E}:\bor{[0,2\pi)\times [0,2\pi)}\to \lhh,
$$
with the property
$$
 \tilde{E}(X\times Y)= E_1(X)\otimes E_2(Y).
$$
Using the function
$$
f:[0,2\pi)\times [0,2\pi) \to [0,2\pi),\ (x,y)\mapsto x-y\
(\textrm{mod } 2\pi)
$$
one gets from $\tilde{E}$ the  observable which is
the difference of the observables $E_1$ and $E_2$:
$$
\Ediff:\bor{[0,2\pi)}\to \lhh,\
\Ediff(X):=
\tilde{E}(f^{-1}(X)).
$$

Using the explicit form \cite{CDLP} of a phase observable
$E:\bo\to\mathcal{L}(\hi)$,
\begin{equation}
E(X)=\sum_{n,m=0}^{\infty} \st{\varphi_n}{\varphi_m}
\frac{1}{2\pi} \int_X e^{i(n-m)\theta}\ud \theta\ \ketbra{n}{m},
\end{equation}
where  $(\varphi_n)_{n\in\N}\subset\hi$ is a sequence of unit
vectors, one easily computes that the difference of $E_1$ and
$E_2$ is
\begin{equation}\label{erotus}
\Ediff(X)=\sum_{n,m,k,l\in\N} \delta_{n-m,l-k}
\st{\varphi^1_n}{\varphi^1_m}\st{\varphi^2_k}{\varphi^2_l}\frac{1}{2\pi}\int_X
e^{i(n-m)\theta}\ud \theta\ \ketbra{n,k}{m,l}.
\end{equation}
Here $\delta$ is the Kronecker delta, $\ketbra{n,k}{m,l}$ stands for the rank one
operator $\hi\otimes\hi\ni\psi\mapsto \st{m,l}\psi|n,k\rangle\in\hi\otimes\hi$, and, for instance,
 $|n,k\rangle$ is the short
hand notation for the tensor product vector $|n\rangle\otimes|k\rangle$.

Let
\begin{eqnarray*}
&& \Sigma N := N\otimes I + I\otimes N\\
&& \Delta\,N := N\otimes I-I\otimes N
\end{eqnarray*}
denote the sum and the difference of the number operators of the two modes,
and let $\Sigma N =\sum_{k\in\N}kP^\Sigma_k$ and $\Delta\,N=\sum_{k\in\Z}kP^\Delta_k$
be their respective spectral decompositions, with the spectral projections
\begin{eqnarray*}
&& P_k^\Sigma=\sum_{n=0}^k\ketbra{k-n,n}{k-n,n},\ k\in\N,\\
&& P_k^\Delta=\sum_{n\geq\max\{0,-k\}}\ketbra{k+n,n}{k+n,n}, \ k\in\Z.
\end{eqnarray*}
Consider the unitary operators
\begin{eqnarray*}
V_{\Sigma}(\alpha)&&=e^{i\alpha\Sigma N},\ \alpha\in\R,\\
V_{\Delta}(\beta)&&=e^{i\beta \Delta\,N},\  \beta\in\R.
\end{eqnarray*}
The difference $\Ediff$ of the phase observables $E_1$ and $E_2$ is \emph{invariant} under $V_{\Sigma}$,
\begin{equation}\label{invarianssi}
V_{\Sigma}(\alpha)\Ediff(X)V_{\Sigma}(\alpha)^*=\Ediff(X),
\end{equation}
for all $\alpha\in\R, X\in\bo$. This condition is equivalent to the commutativity
of $\Sigma N$ and $\Ediff$, that is,
$$
P_k^\Sigma\Ediff(X)=\Ediff(X)P_k^\Sigma
$$
for all $k\in\N, X\in\bo$.
Since the number sum is a projection valued observable $k\mapsto P_k^\Sigma$, the commutativity
of $\Sigma N$ and $\Ediff$ equals with their being (functionally) coexistent, that is, they have a joint
observable, see, for instance,  \cite{LaPul}.
%
%
It is another immediate observation that $\Ediff$ satifies the following \emph{covariance} condition under $V_{\Delta}$:
\begin{equation}\label{kovarianssi}
V_{\Delta}(\beta)\Ediff(X)V_{\Delta}(\beta)^*= \Ediff(X\dotplus 2\beta),
\end{equation}
for all $\beta\in\R, X\in\bo$.

Let
$$
\Theta(\alpha,\beta)=e^{i\alpha N\otimes I + i\beta I\otimes N},\ \alpha,\beta \in \R.
$$
Since
$$
\Theta(\alpha,\beta)=V_{\Sigma}(\frac{\alpha}{2})V_{\Delta}(\frac{\alpha}{2})V_{\Sigma}(\frac{\beta}{2})V_{\Delta}(-\frac{\beta}{2}),
$$
we observe that the invariance and covariance conditions
(\ref{invarianssi}) and (\ref{kovarianssi}) are equivalent with
the condition
\begin{equation}\label{ehto1.0}
\Theta(\alpha,\beta)\Ediff(X)\Theta(\alpha,\beta)^*=\Ediff(X\dotplus (\alpha - \beta))
\end{equation}
for all $X\in \bor{[0,2\pi)}$ and $\alpha,\beta\in\R$.

These observations lead us to the following definition.

\begin{definition}\label{def1}
A \emph{phase difference observable} is
a normalized positive operator measure $E:\bo\to\lhh$
which satisfies the covariance condition
\begin{equation}\label{ehto1}
\Theta(\alpha,\beta) E(X)\Theta(\alpha,\beta)^*= E(X\dotplus
(\alpha - \beta))
\end{equation}
for all $X\in \bor{[0,2\pi)}$ and $\alpha,\beta\in\R$.
\end{definition}

In the next sections
we characterize all phase difference
observables and we also give a  necessary and sufficient
condition for a  phase difference observable
to be a difference of  two (one-mode) phase observables.

\section{Direct method}\label{direct}

The following lemma simplifies the proof of the Theorem 1 below.
\begin{lemma} \label{joukko}
Let $q\in\Z$ and let $\nu_q:\bo\to\C$ be a $\sigma$-additive set function.
Then $\nu_q(X\dotplus\theta)=e^{iq\theta}\nu_q(X)$ for all
$X\in\mathcal B([0,2\pi))$ and $\theta\in[0,2\pi)$ if and only if
$\nu_q(X)=c_q \frac{1}{2\pi}\int_Xe^{i q\theta}\mathrm d\theta$ for all
$X\in\bo$, where $c_q\in\C$.
\end{lemma}

\begin{proof}
The 'if'-part of the lemma is clear, so we have to prove 'only if'-statement.
Assume  that $\nu_q(X\dotplus\theta)=e^{iq\theta}\nu_q(X)$ for all
$X\in\mathcal B([0,2\pi))$ and $\theta\in[0,2\pi)$. Since $[0,2\pi)\dotplus\theta=[0,2\pi)$ it follows that
$$
\nu_q([0,2\pi))=\nu_q([0,2\pi)\dotplus\theta)=e^{iq\theta}\nu_q([0,2\pi))
$$
and thus $\nu_q([0,2\pi))=c_0\delta_{0,q}$, where $c_0$ is a complex
constant. The rest of the proof is same as the proof of Lemma 1 in
\cite{CDLP}.
\end{proof}

Let
$E:\bo\to\mathcal{L}(\hi\otimes\hi)$ be an arbitrary operator measure, that is, an
$\mathcal{L}(\hi\otimes\hi)$-valued map defined on $\bo$
which is
$\sigma$-additive with respect to the weak operator topology.

\begin{theorem}\label{theorem1}
\begin{itemize}
\item[a)]
If the  operator measure  $E:\bo\to\lhh$ satisfies the covariance
condition $(\ref{ehto1})$, then for all $X\in\bo$,
\begin{equation}\label{z}
E(X)= \sum_{n,m,k,l=0}^{\infty}
c_{n,m,k,l}\frac{1}{2\pi}\int_Xe^{i(n-m)\theta}\mathrm{d}\theta\
|n,k\rangle\langle m,l|,
\end{equation}
where $c_{n,m,k,l}\in\C$, and
$c_{n,m,k,l}=0$ if $n-m\ne l-k$,
for all $n,m,k,l\in\N$.
\item[b)] If, in addition,  $E$ is positive, that is, $E(X)\geq O$ for all $X\in\bo$, then
$\sum_{n,m,k,l=0}^Nc_{n,m,k,l}\ketbra{n,k}{m,l}\geq O$ for all
$N\in\mathbb N$, and
\item[c)] if $E$ is normalized, that is, $E([0,2\pi))=I$, then
  $c_{n,n,k,k}=1$ for all $n,k\in\mathbb N$.
\end{itemize}
\end{theorem}

\begin{proof}
\begin{itemize}
\item[a)]Using the covariance condition we get
$$
\langle n,k| E(X\dotplus (\alpha -\beta)) |m,l\rangle = e^{i\alpha
  (n-m)+i\beta (k-l)} \langle n,k|E(X)|m,l\rangle
$$
for all $n,m,k,l\in\N$, $\alpha,\beta\in\R$ and $X\in\bo$.
Choosing $\alpha=\beta$ it follows that $\langle
n,k|E(X)|m,l\rangle =0$ if $n-m\neq l-k$. Denote
$$
\nu_{n,m,k} (X) := \langle n,k | E(X)| m,k+n-m\rangle.
$$
We get
\begin{eqnarray*}
\nu_{n,m,k}(X\dotplus (\alpha-\beta)) &=& \langle n,k | E(X\dotplus (\alpha-\beta))| m,k+n-m\rangle \\
&=& e^{i(\alpha(n-m)+\beta (m-n))}\langle n,k |E(X)|m,k+n-m\rangle\\
&=& e^{i(\alpha-\beta)(n-m)} \nu_{n,m,k}(X).
\end{eqnarray*}
Taking $q=n-m$, Lemma \ref{joukko} now gives
Equation (\ref{z}).

\item[b)] If $\sum_{n,m,k,l=0}^N c_{n,m,k,l}\langle\psi|{n,k}\rangle\langle{m,l}|\psi\rangle<0$
for some $N\in\N$ and $\psi\in\hi\otimes\hi$ then, due to the continuity of the density function,
one may choose an $\epsilon>0$ such that
$\langle P_N\psi|E([0,\epsilon))P_N\psi\rangle < 0$, where
$P_N:=\sum_{n,k=0}^N|n,k\rangle\langle n,k|$. This is a contradiction.

\item[c)] This is a direct check.
\end{itemize}
\end{proof}

To prove the converse of the above theorem, consider a positive normalized operator
measure $\tilde{E}:\bor{[0,2\pi)\times [0,2\pi)}\to \lhh$ and a set of complex numbers
$\tilde{\bold c} :=(\tilde c_{n,m,k,l})_{n,m,k,l\in\mathbb N}$.
We say that $\tilde E$ is $\Theta$-{\em covariant} if
$$
\Theta(\alpha,\beta)\tilde E(Z)\Theta(\alpha,\beta)^*
= \tilde E(Z\dotplus(\alpha,\beta))
$$
for all $Z\in \bor{[0,2\pi)\times [0,2\pi)}$, $\alpha,\beta\in\mathbb R$, with $\dotplus$
meaning (componentwise) addition mod $2\pi$, and we say that
$\tilde{\bold c}$ is {\em normalized} {\em positive semidefinite} if
\begin{eqnarray*}
&&\ \ \ \ \tilde c_{n,n,m,m} = 1,\\
&&\sum_{n,m,k,l=0}^N\tilde c_{n,m,k,l}\ketbra{n,k}{m,l}\geq O,
\end{eqnarray*}
for all $n,m,N\in\mathbb N$.

With the above notations the following theorem is then obtained.
Its proof is essentially the same as in the one-dimensional case \cite{LaPe2}
so that we omit it here.

\begin{theorem}\label{thm2}
\begin{itemize}
\item[a)]
If $\tilde E$ is $\Theta$-covariant, then there is a normalized positive semidefinite $\tilde{\bold c}$
such that for any $Z\in\bor{[0,2\pi)\times [0,2\pi)}$,
\begin{equation}\label{kaava1}
\tilde E(Z) = \sum_{n,m,k,l=0}^\infty \tilde c_{n,m,k,l}
\int_Ze^{i[(n-m)x+(k-l)y]}\textstyle{\frac{dx}{2\pi}\frac{dy}{2\pi}}\ketbra{n,k}{m,l}.
\end{equation}
\item[b)] If $\tilde{\bold c}$ is normalized positive semidefinite, then  formula
$(\ref{kaava1})$ defines (weakly) a $\Theta$-covariant normalized positive operator
measure $\tilde{E}:\bor{[0,2\pi)\times [0,2\pi)}\to \lhh$.
\end{itemize}
\end{theorem}

By the above theorem, given a normalized positive semidefinite set
of complex numbers $\tilde{\bold c}$ we
get a $\Theta$-covariant normalized positive operator measure $\tilde{E}$.
Consider again the function $f(x,y) =x-y\
(\textrm{mod } 2\pi)$, defined on the rectangle $[0,2\pi)\times [0,2\pi)$.
Then the map
$\bo\ni X\mapsto \tilde E^f(X):=\tilde E(f^{-1}(X))\in \lhh$
is a phase difference observable with the structure given in Equation (\ref{z}), where now
\begin{equation}\label{c}
c_{n,m,k,l}=\tilde c_{n,m,k,l}\delta_{n-m,l-k}.
\end{equation}

\begin{remark}\label{vaihtoehto}
Let $(\psi_{n,k})_{n,k\in\N}$ be a set of unit vectors in
$\hi$. It is clear that defining
\begin{equation}\label{form1}
\tilde{c}_{n,m,k,l}=\st{\psi_{n,k}}{\psi_{m,l}},
\end{equation}
$\tilde{\bold c}$ is  normalized positive semidefinite. Also the
converse is true, any normalized positive semidefinite set of
complex numbers is of the form (\ref{form1}). Construction of a set of
unit vectors $\{\psi_{n,k}\}_{n,k\in\N}$ for a given $\tilde{\bold c}$
is similar than the one given in \cite{CDLP}, section II.B.
We note also that if $\tilde{\bold c}$ is positive semidefinite defined by
unit vectors $(\psi_{n,k})_{n,k\in\N}$ then  $\bold c$ defined as
in Eq.\ (\ref{c}) is positive semidefinite since one may choose a sequence
$(\psi_{n,k}\otimes|n+k\rangle)_{n,k\in\N}$ of unit vectors to define $\bold c$.
\end{remark}

\begin{remark}
Equation (\ref{c}) shows that $\Theta$-covariant observables
$\tilde E$ form a "wider" class of observables than phase
difference observables of Definition~\ref{def1} in the sense that
there are many $\Theta$-covariant observables which give the same
phase difference observable, and any phase difference observable
with $\bold c$ defines a $\Theta$-covariant observable which has,
for instance, the same $\bold c$ as its structure unit.
One may define an equivalence relation between $\Theta$-covariant
observables as follows: two $\Theta$-covariant observables with
$\tilde{\bold c}$ and $\tilde{\bold d}$ are equivalent if $\tilde
c_{n,m,k,l}=\tilde d_{n,m,k,l}$ for all $n,m,k,l\in\N$, $n-m=l-k$,
that is, if they define the same phase difference observable.
\end{remark}

\begin{remark}
Using (\ref{z}), it easy to see that any phase difference
observable $E$ has a uniform distribution in states where one mode
is in a number state. For example, if
$\psi:=\varphi\otimes|s\rangle$, $\varphi\in\hi$, $\|\psi\|=1$,
$s\in\N$, then
$$
\langle\psi|E(X)|\psi\rangle = \frac{1}{2\pi}\int_X \ud
\theta,\quad n,k\in\N,\;X\in\bo.
$$
Moreover, one may  also witness  that there is no projection valued phase
difference observable. For example,
$$
\langle 0,0|E(X)^2|0,0\rangle = \left|\frac{1}{2\pi}\int_X
  \ud\theta \right|^2
$$
and choosing $X=[0,\pi)$ we get $\langle
0,0|E([0,\pi))^2|0,0\rangle=\frac{1}{4}$. Compared to $\langle
0,0|E([0,\pi))|0,0\rangle=\frac{1}{2}$, this shows that a phase
difference observable cannot be a spectral measure.
\end{remark}

\section{Group theoretical solution}\label{covdis}

In \cite{CDLP} all phase observables were calculated using a
generalized imprimitivity theorem due to Cattaneo \cite{Cattaneo}.
Here we follow the same method to give an alternative way to
derive the structure of phase difference observables. In using
group theoretical methods, it is convenient to work in the torus
$\mathbb T:=\{z\in\C\,:\,|z|=1\}$, instead of phase interval
$[0,2\pi)$ where addition is to be taken  modulo $2\pi$. We regard
$\mathbb T$  as a compact (second countable) Abelian group and we
let  $\mu$ denote its  Haar measure. The product group
$\torus\times\torus$ has a unitary representation
 $U$ on $\ltok$, defined by
\begin{equation}\label{tesi1}
[U(a,b)f](z_1,z_2)=f(az_1,bz_2).
\end{equation}
To solve the covariance condition (\ref{ehto1}), we will first
characterize all positive normalized operator measures
$F:\boto\to\mathcal{L}(\ltok)$ that satisfy
\begin{equation}\label{ehto:tt}
U(a,b)F(X)U(a,b)^*=F(ab^{-1}X)
\end{equation}
for all $X\in\boto$, $a,b\in\torus$. The canonical spectral
measure $F_{\rm can}$ satisfying this condition is of the form
\begin{equation}\label{cano}
[F_{\rm can}(X)f](z_1,z_2)=\chi_X(z_1^{-1}z_2)f(z_1,z_2),
\end{equation}
where $\chi_X$ is the characteristic function of the set $X$.

Notice that $U(a,b)=U(a,1)U(1,b)$, so that the covariance conditions
\begin{equation}\label{ehto:t1}
U(a,1)F(X)U(a,1)^*=F(aX),\quad a\in\torus,X\in\boto,
\end{equation}
and
\begin{equation}\label{ehto:t2}
U(1,b^{-1})F(X)U(1,b^{-1})^*=F(bX),\quad b\in\torus,X\in\boto,
\end{equation}
taken together are equivalent with the  condition (\ref{ehto:tt}).
We will denote the representation $a\mapsto U(a,1)$ as $U_1$ and the
representation $a\mapsto U(1,a^{-1})$ as $U_2$.

Covariance condition (\ref{ehto:tt}) can be solved by looking the
action $z\mapsto ab^{-1}z$ of $\torus\times\torus$ on $\torus$ and
noting that the stability subgroup is $\torus$ \cite{italy}.
Here we proceed in a different way. We characterize the normalized positive
operator measures satisfying  separately conditions (\ref{ehto:t1})
and (\ref{ehto:t2}). Then  we combine the results to obtain
operator measures satisfying condition (\ref{ehto:tt}). Finally,
we go back  to the original Hilbert space $\hi$ and to the phase  interval $[0,2\pi)$
to get all the phase difference observables.

Let  $F$ be a normalized positive operator measure satisfying
condition (\ref{ehto:t1}). Since the action $z\mapsto az$ of
$\torus$ on itself is transitive, $(U_1,F)$ is a transitive system
of covariance based on $\torus$ and, hence, $(U_1,F)$ is described
by \cite[Proposition~2]{Cattaneo}. In order to apply the cited
result, let us notice the following facts. The stability subgroup
of any point of $\torus$ is the trivial subgroup $\{1\}$.  The
trivial representation $\sigma$ of $\{1\}$ acting on  $\lto$
contains all the (trivial) representations of $\{1\}$ and the
corresponding imprimitivity system $(R,P)$  for $\torus$ based on
$\torus$ induced by $\sigma$ acts on  $L^2(\torus,\mu,\lto)\simeq
\ltok$ as
\begin{eqnarray}
(R(a)\fii)(z_1,z_2) & = & \fii(a^{-1}z_1,z_2), \label{ind1}\\
(P(X)\fii)(z_1,z_2) & = & \chi_X(z_1)\fii(z_1,z_2), \label{ind2}
\end{eqnarray}
where $\fii\in\ltok, a\in\torus, X\in\boto$ and
$z_1,z_2\in\torus$.

Proposition 2 of \cite{Cattaneo} shows that, given a normalized
covariant positive operator measure $F$, there exists an isometry
$$W_1:\ltok\to\ltok,$$
which intertwines the action $U_1$ with $R$ and such that
\begin{equation}\label{correspondence}
F(X)=W_1^*P(X)W_1,\ \ \ \ X\in\boto.
\end{equation}
Conversely, given an intertwining isometry $W_1$ from $\ltok$ to
$\ltok$, equation (\ref{correspondence}) defines a  positive normalized
operator measure $F$ satisfying equation (\ref{ehto:t1}).

Hence, to classify all normalized positive operator measures
satisfying condition (\ref{ehto:t1}), one has to determine all the
isometric mappings $W_1$ such that
\begin{equation}\label{intertwine}
W_1U_1(a)=R(a)W_1,\ \ \ a\in\torus.
\end{equation}

To perform this task, observe that the monomials  $e_n$, $n\in\Z$,
$e_n(z)=z^n$, $z\in\torus$, form an orthonormal basis of $\lto$.
Similarly, the product vectors
$$
(e_n e_k)(z_1,z_2)=e_n(z_1)e_k(z_2)=z_1^n z_2^k,\  n,k\in\Z,
$$
 form an orthonormal basis of $\ltok$.

The action of $U_1$ in this base is
$$
U_1(a)(e_ne_k)=a^{n}(e_ne_k)
$$
and the action of $R$ is simply
$$
R(a)(e_ne_k)=a^{-n}(e_ne_k).
$$
From equation (\ref{intertwine}) we get
\begin{equation*}
R(a)W_1(e_ne_k) = W_1U_1(a)(e_ne_k) = a^{n}W_1(e_ne_k)
\end{equation*}
for all $n,k\in \Z$.
It follows that $W_1(e_ne_k)$ must be in the vector space
$\overline{\mathrm{span}}\{(e_{-n}e_j)\}_{j\in\Z} \simeq \lto$. This
means that $W_1(e_ne_k)=(e_{-n}\psi_{n,k})$, where $\psi_{n,k}$ is
some unit vector in $\lto$.

The matrix elements of $F$ in the basis $\{(e_ne_k)\}_{n,k\in\mathbb N}$ are thus:
\begin{eqnarray}\label{matrix1}
\langle (e_ne_k)|F(X)(e_me_l)\rangle &=&
\langle (e_ne_k)|W_1^*P(X)W_1(e_me_l)\rangle\\
&=&
\langle W_1(e_ne_k)|P(X)W_1(e_me_l)\rangle \nonumber \\
&=&
\langle (e_{-n}\psi_{n,k})|P(X)(e_{-m}\psi_{m,l})\rangle \nonumber\\
&=& \langle \psi_{n,k}|\psi_{m,l}\rangle\,\int_Xz^{n-m}\,\mathrm d\mu(z).\nonumber
\end{eqnarray}

We consider next  condition (\ref{ehto:t2}). Like in the
previous case, $U_2$ and a normalized positive operator measure
$F$ satisfying (\ref{ehto:t2}), form a transitive system of
covariance based on $\torus$. The corresponding imprimitivity
system is  the same pair $(R,P)$, defined in Equations (\ref{ind1})
and (\ref{ind2}). The action of $U_2$ in the basis $\{(e_ne_k)\}_{n,k\in\mathbb N}$ is
$$
U_2(a)(e_ne_k)=a^{-k}(e_ne_k).
$$
If $W_2$ is an isometry intertwining representations $U_2$ and $R$, then
\begin{equation}\label{inter2}
R(a)W_2(e_ne_k)=W_2U_2(a)(e_ne_k)=a^{-k}W_2(e_ne_k).
\end{equation}
Thus $W_2(e_ne_k)$ must be in the vector space
$\overline{\mathrm{span}}\{(e_ke_j)\}_{j\in\Z} \simeq \lto$ and
$W_2(e_ne_k)=e_k\varphi_{n,k}$ for some unit vector
$\varphi_{n,k}\in\lto$.

Matrix elements are now:
\begin{eqnarray}\label{matrix2}
\langle (e_ne_k)|F(X)(e_me_l)\rangle &=&
\langle (e_ne_k)|W_2^*P(X)W_2(e_me_l)\rangle\\
&=&
\langle W_2(e_ne_k)|P(X)W_2(e_me_l)\rangle\nonumber\\
&=&
\langle (e_k\varphi_{n,k})|P(X)(e_l\varphi_{m,l})\rangle\nonumber\\
&=& \langle \varphi_{n,k}|\varphi_{m,l}\rangle\,\int_X z^{l-k}\,\mathrm d\mu(z).\nonumber
\end{eqnarray}

Assume now that $F$ is a normalized positive operator measure that
satisfy condition (\ref{ehto:tt}), or,
equivalently,  conditions (\ref{ehto:t1}) and (\ref{ehto:t2}). This
means that the matrix elements
(\ref{matrix1}) and (\ref{matrix2}) are the same:
\begin{equation}
\langle \psi_{n,k}|\psi_{m,l}\rangle\,\int_X z^{n-m}\,\mathrm d\mu(z)=
\langle \varphi_{n,k}|\varphi_{m,l}\rangle\,\int_X z^{l-k}\,\mathrm
d\mu(z)
\end{equation}
for all $n,m,k,l\in\Z$ and $X\in\boto$. From this we get $n-m=l-k$ and
$ \langle \psi_{n,k}|\psi_{m,l}\rangle=\langle \varphi_{n,k}|\varphi_{m,l}\rangle.$

We summarize the above construction in the following theorem.

\begin{theorem}\label{the:tt}
Any normalized positive operator measure $F:\boto\to\ltok$
satisfying covariance condition $(\ref{ehto:tt})$ is of the form
\begin{equation}\label{e:tt}
F(X) = \sum_{n,m,k,l \in \Z} \delta_{n-m,l-k} \st{\psi_{n,k}}{\psi_{m,l}}
\int_X z^{n-m}\ud \mu(z)\ \ketbra{e_ne_k}{e_me_l}
\end{equation}
for some set $(\psi_{n,k})_{n,k\in\Z}\subset\lto$ of unit vectors.
\end{theorem}

We note that in  (\ref{e:tt}) only the inner products of the
vectors $\psi_{n,k}$ are relevant. Thus two set of unit vectors
$(\psi_{n,k})_{n,k\in\Z}$ and $(\eta_{n,k})_{n,k\in\Z}$ define the
same positive operator measure exactly when
\begin{equation*}
\delta_{n-m,l-k}\st{\psi_{n,k}}{\psi_{m,l}}=\delta_{n-m,l-k}\st{\eta_{n,k}}{\eta_{m,l}}
\end{equation*}
for all $n,m,k,l\in\Z$.

\begin{example}
The canonical spectral measure $F_{\rm can}$ of Equation
(\ref{cano}) written in the above form is simply
\begin{equation*}
F_{\rm can}(X)= \sum_{n,m,k,l \in \Z} \delta_{n-m,l-k}\int_X z^{n-m}\ud
\mu(z)\ \ketbra{e_ne_k}{e_me_l},
\end{equation*}
showing that $F_{\rm can}$ can be defined by a set
$(\psi_{n,k})_{n,k\in\Z}$, where $\psi_{n,k}=\psi$ for all $n,k\in\Z$ and $\psi$ is any unit vector.
\end{example}

We are now ready to solve the covariance condition (\ref{ehto1}).
Let $\hi$ be a complex separable Hilbert space with an orthonormal
basis $\{|n\rangle\}_{n\in\N}$, and $T:\hi \otimes \hi\to\ltok$ be
a linear isometry with the property
$$
T|n,m\rangle = e_ne_m,\ {\rm for\ all }\ n,m\in\N.
$$
If $[0,2\pi)$ is identified with $\torus$ by the mapping
$\alpha\mapsto e^{i\alpha}$, then $\Theta$ can be regarded as
a unitary representation of $\torus\times\torus$.
Clearly, $T$ intertwines representations $\Theta$ and $U$, $T\Theta=UT$.
If $\tilde{F}:\boto\to\mathcal{L}(\hi\otimes\hi)$ satisfies  the equation
\begin{equation}\label{ehto:vali}
\Theta(a,b)\tilde{F}(X)\Theta(a,b)^*=\tilde{F}(ab^{-1}X)
\end{equation}
for all $a,b\in\torus,\ X\in\boto$, then $F(X):=T\tilde{F}(X)T^*$
is a normalized positive operator measure having  property
(\ref{ehto:tt}). Moreover, if $F:\boto\to\mathcal{L}(\ltok)$
satisfies  condition (\ref{ehto:tt}), then $X\to T^*F(X)T$ is a
normalized positive operator measure acting in
$\mathcal{L}(\hi\otimes\hi)$ and satisfying (\ref{ehto:vali}).
Using theorem \ref{the:tt}, one thus has the following result.

\begin{theorem}\label{cov:sol}
A normalized positive operator measure $E:\bo\to\mathcal{L}(\hi
\otimes \hi)$ is a phase difference observable if and only if
\begin{equation}\label{e:sol}
E(X) = \sum_{n,m,k,l\in\N}
\delta_{n-m,l-k}\st{\xi_{n,k}}{\xi_{m,l}}\frac{1}{2\pi}\int_X
e^{i(n-m)\theta}\ud\theta\ \ketbra{n,k}{m,l},
\end{equation}
for some set of unit vectors $(\xi_{n,k})_{n,k\in\N}$ of $\hi$.
\end{theorem}

In view of  Remark \ref{vaihtoehto}, this result is the same  as the one
obtained in
Section \ref{direct}. Since $T:\hi\otimes\hi\to\ltok$ is not
surjective, there is no projection valued phase difference observable.

\begin{remark}
The moment operators $E^{(r)},r\in\N$, of the phase difference
observable $E$ are defined as
$$
E^{(r)}:=\int_0^{2\pi}\theta^r\ud E(\theta)
$$
and they are bounded self-adjoint operators. By direct
calculation we get
$$
\st{n,k}{E^{(1)}|m,l}=\delta_{n-m,l-k}\st{\xi_{n,k}}{\xi_{m,l}}\frac{i}{m-n}
$$
when $n\neq m$.
For $n=m$ one gets
$$
\st{n,k}{E^{(1)}|n,l}=\pi \delta_{k,l}.
$$
Thus the phase difference observable $E$ is uniquely determined by its
first moment operator $E^{(1)}$. This is notable since $E$ is not
projection valued. The same result holds also for phase observables, see \cite{DHLP} for a further discussion of this
conundrum.

Similarly, the \emph{r}th cyclic moment operator of $E$ is defined as
the operator $C_{E}^{(r)}$,
$$
C_{E}^{(r)}:=\int_0^{2\pi}e^{ir\theta}\ud
E(\theta),\quad r\in\N.
$$
They are easily determined to be
$$
C_{E}^{(r)}=\sum_{n,l=0}^{\infty} \st{\xi_{n,l+r}}{\xi_{n+r,l}}\
\ketbra{n,l+r}{n+r,l}.
$$

Since $C_E^{(1)}|0,0\rangle=0$, the first cyclic moment is not
unitary. This is another way to see the already mentioned fact that
there is no projection valued phase difference observable.

\end{remark}

\section{Phase difference observable {\em vs.}
difference of phase observables}\label{diffe}

%

Till now we have characterized in two different ways the phase difference observables,
and we have also constructed explicitly the difference of two phase observables.
The following proposition characterizes those phase difference
observables which are, that is, can be expressed as, the difference of
two phase observables. It's proof is a direct comparison of formulas
(\ref{erotus}) and (\ref{e:sol}).

\begin{proposition}\label{erotus-erotus}
Let $E:\bo\to\mathcal{L}(\hi\otimes\hi)$ be a phase difference
observable, characterized by a set $(\xi_{n,k})_{n,k\in\N}$. Observable
E is a difference  of two phase observables if and
only if there are sequences $(\varphi^1_n)_{n\in\N}$ and
$(\varphi^2_n)_{n\in\N}$ of unit vectors in $\hi$ such that
\begin{equation} \label{si1}
\delta_{n-m,l-k}\st{\xi_{n,k}}{\xi_{m,l}}=\delta_{n-m,l-k}\st{\varphi^1_n}{\varphi^1_m}\st{\varphi^2_k}{\varphi^2_l}
\end{equation}
for all $n,k,m,l\in\N$.
\end{proposition}

The next example shows that there are phase difference observables that
are not difference of two phase observables. It also opens the question of 
finding physically meaningful conditions for  Proposition~\ref{erotus-erotus}.

\begin{example}
Fix an arbitrary unit vector $\psi\in\hi$ and let $\theta_j,j=1,2,3,4,$
be real numbers. Define $\xi_{0,2}=e^{i\theta_1}\psi$, $\xi_{2,2}=e^{i\theta_2}\psi$, $\xi_{0,4}=e^{i\theta_3}\psi$, $\xi_{2,4}=e^{i\theta_4}\psi$ and
 $\xi_{n,k}=\psi$ otherwise. Assume now that there are sequences
 $(\varphi^1_n)_{n\in\N}$ and $(\varphi^2_n)_{n\in\N}$ such that
 equation (\ref{si1}) holds. Then
\begin{eqnarray*}
e^{i\theta_4} &=& \st{\xi_{3,3}}{\xi_{2,4}} =
\st{\varphi^1_3}{\varphi^1_2}\st{\varphi^2_3}{\varphi^2_4} \\
&=&
\frac{\st{\varphi^1_3}{\varphi^1_2}\st{\varphi^2_1}{\varphi^2_2}\st{\varphi^1_1}
{\varphi^1_0}\st{\varphi^2_3}{\varphi^2_4}}{\st{\varphi^1_1}{\varphi^1_0}\st{\varphi^2_1}{\varphi^2_2}}\\
&=&
\frac{\st{\xi_{3,1}}{\xi_{2,2}}\st{\xi_{1,3}}{\xi_{0,4}}}{\st{\xi_{1,1}}{\xi_{0,2}}} = e^{i(\theta_2+\theta_3-\theta_1)}.
\end{eqnarray*}
Choosing the numbers $\theta_j$ in such a way that $e^{i\theta_4}\neq
e^{i(\theta_2+\theta_3-\theta_1)}$ we thus get a contradiction.
\end{example}

From equation (\ref{si1}) it is also clear that
two different pairs of phase observables may define the same phase
difference observable. 



\

We close this section with a terminological choice.
We say that a phase difference observable is 
{\em canonical} if it is  the difference  of two canonical phase observables
and we denote it  by $\Ecan$.
Since the canonical phase observable $\Ec$
has the structure
$$
\Ec(X)=\sum_{n\in\N} \frac{1}{2\pi} \int_X e^{i(n-m)\theta}\ud\theta\
\ketbra{n}{m},
$$
the  explicit form of $\Ecan$ can be read from both  (\ref{erotus}) and
(\ref{e:tt}) with the involved inner products equal to one in each
case. Some properties of canonical phase difference observable are
discussed in sections 6 and 8.

\section{Radon-Nikod\'ym derivatives and the phase difference representation}
Let $T$ be a state (positive trace-one operator) on
$\hi\otimes\hi$, let $E$ be a phase difference observable with
$\bold c$, and let $\tilde E$ be a $\Theta$-covariant observable
with $\tilde{\bold c}$. Using similiar methods as in \cite[Sec.\
V]{Pell2}, one can show that
\begin{eqnarray*}
{\rm tr}(TE(X))&=&\frac{1}{2\pi}\int_X g^E_T(\theta)\,\mathrm d\theta,
\;\;\;\;X\in\mathcal B\left([0,2\pi)\right),\\
{\rm tr}(T\tilde E(Z))&=&\frac{1}{(2\pi)^2}\int_Z\tilde g^{\tilde E}_T(x,y)
\,\mathrm dx\,\mathrm dy,
\;\;\;\;Z\in\mathcal B\left([0,2\pi)\times[0,2\pi)\right),
\end{eqnarray*}
where
\begin{eqnarray*}
g^E_T(\theta)&=&\sum_{n,m,k,l=0}^\infty c_{n,m,k,l}e^{i(n-m)\theta}\langle m,l|T|n,k\rangle,\\
\tilde g^{\tilde E}_T(x,y)&=&\sum_{n,m,k,l=0}^\infty\tilde c_{n,m,k,l}
e^{i(n-m)x+i(k-l)y}\langle m,l|T|n,k\rangle
\end{eqnarray*}
for $\mathrm d\theta$-almost all $\theta\in\R$ and
for $\mathrm dx\mathrm dy$-almost all $(x,y)\in\R^2$.
The above notations $\sum_{n,m,k,l=0}^\infty$ mean
that for some increasing subsequences
$(s_t)_{t\in\mathbb N}\subseteq\mathbb N$,
$\sum_{n,m,k,l=0}^\infty=\lim_{t\to\infty}\sum_{n,m,k,l=0}^{s_t}$.
It is easy to see that if $E$ is constructed from $\tilde E$ (that is,
Eq.\ (\ref{c}) holds) then
$$
g^E_T(\theta)=\frac{1}{2\pi}\int_0^{2\pi}\tilde g^{\tilde E}_T(x+\theta,x)\mathrm dx.
$$
Since $\hi$ is isomorphic with the Hardy space $H^2$ of the unit circle
spanned by the vectors $e_n$, $n\in\N$, one can consider any $\psi\in\hi$
as an element of $H^2$, that is, as a function (or equivalence class of functions).
Using this interpretation, for any $\varphi,\psi\in\hi$ and $X\in\bo$, one may write
\begin{equation}\label{dis}
\langle \varphi\otimes\psi| \Ecan(X)\varphi\otimes\psi\rangle =
\frac{1}{2\pi}\int_X\frac{1}{2\pi}\int_{0}^{2\pi}|\varphi(x+\theta)|^2|\psi(x)|^2\ud x\ud\theta.
\end{equation}
This phase difference distribution was
first suggested by Barnett and Pegg \cite{BP90,PB97}.


\section{Classical limit}
Like in the one-mode case \cite{PBCov},
it is easy to show that for any operator measure $E:\bo\to\lhh$
the condition
$$
\left\langle z_1e^{i\alpha},z_2e^{i\beta}\right|
E(X)\left|z_1e^{i\alpha},z_2e^{i\beta}\right\rangle
=\left\langle z_1,z_2\right|
E(X\dotplus(\alpha - \beta))\left|z_1,z_2\right\rangle,
$$
$z_1,z_2\in\mathbb C$, $\alpha,\beta\in\mathbb R$, $X\in
\bor{[0,2\pi)}$, equals the covariance condition (\ref{ehto1})
where $|z_1,z_2\rangle:=|z_1\rangle\otimes|z_2\rangle$ is a
two-mode coherent state.

Suppose that $E^{\rm diff}$ is the difference of phase observables $E_1$
and $E_2$ with $(c^1_{n,m})$ and $(c^2_{n,m})$, respectively.
If, for example, $\lim_{n\to\infty}c^2_{n,n+k}=e^{ik\alpha}$ for all $k\ge1$,
$\alpha\in[0,2\pi)$, then
for any continuos function $g:\,[0,2\pi]\to\C$
$$
\lim_{|z|\to\infty \atop
\arg z\;{\rm fixed}}\int_0^{2\pi}g(x)\mathrm{d}\langle z|E_2(x)|z\rangle=
g(\arg z-\alpha)
$$
(see, \cite[Th.\ 7.1]{JPPL00}).
Let $g^{E_n}_{|z\rangle}:\,[0,2\pi]\to[0,\infty)$ be a continuos Radon-Nikod\'ym derivative
of the probability measure $X\mapsto\langle z|E_n(X)|z\rangle$, $n=1,2$.
Then
$$
\lim_{|z_2|\to\infty \atop \arg z_2\;{\rm fixed}}
\frac{1}{2\pi}\int_0^{2\pi}g^{E_1}_{|z_1\rangle}(x+\theta)
g^{E_2}_{|z_2\rangle}(x)\mathrm{d}x=g^{E_1}_{|z_1\rangle}(\theta+\arg z_2-\alpha)
$$
which implies the following proposition:
\begin{proposition}
For any $X\in\mathcal B([0,2\pi))$,
$$
\lim_{|z_2|\to\infty \atop \arg z_2\;{\rm fixed}}
\langle z_1,z_2|E^{\rm diff}(X)|z_1,z_2\rangle
=\left\langle z_1\right|E_1(X+\arg z_2-\alpha)\left|z_1\right\rangle.
$$
\end{proposition}
This means that, in the classical limit $|z_2|\to\infty$ of the second mode,
the two-mode theory reduces to a single-mode theory.
Moreover, if also $\lim_{n\to\infty}c^1_{n,n+k}=e^{ik\alpha'}$ for all
$k\ge1$, then
$$
\lim_{|z_1|,|z_2|\to\infty \atop \arg z_1,\,\arg z_2\;{\rm fixed}}
\langle z_1,z_2|E^{\rm diff}(X)|z_1,z_2\rangle
=\delta_{\arg z_1-\arg z_2-\alpha'+\alpha}(X)
$$
where $\delta_p$ is a Dirac measure concentrated on the point $p$.
This is the classical limit of the two-mode system.

\begin{remark}\label{rem}\rm
It is known from the theory of homodyne detection
\cite{MQSL,BaT48} that when the reference mode is in a large
amplitude coherent state $|z\rangle$, $|z|\gg0$, the lowering
operator $a$ of the reference mode can be replaced with the
"classical" observable $z\,I$ in practical calculations. This
means that the energy and the phase of the reference field are well
known and fixed. A similar result also holds for the difference of
phase obsevables, as well.

Let $\alpha\in[0,2\pi)$ and define a {\it fixed-phase observable}
$$
F_\alpha:\,\bo\to\lh,\;X\mapsto\delta_{\alpha}(X)\,I
$$
where $\delta_\alpha$ is the Dirac measure concentrated on $\alpha$.
The fixed-phase observable $F_\alpha$ is the spectral measure of a self-adjoint
operator $\alpha\,I$ and, thus, it is not a phase observable.
If we choose the phase observable $E_2$ to be the
fixed-phase observable $F_\alpha$ (this can be done similarly as in the case
of two phase observables although $F_\alpha$ is not
covariant), then
$E^{\rm diff}(X)=E_1(X\dot+\,\alpha)\otimes I$, that is,
the "phase difference" $E^{\rm diff}$ and the single-mode phase $E_1$
are practically the same observables (up to unitary equivalence or the choice
of the reference phase $\alpha$).
\end{remark}

\section{Ban's theory}
In the series of papers \cite{Ban1,Ban2,Ban3} Ban has proposed a
unitary two-mode phase operator in relation to the number difference.
To discuss Ban's theory in the
present context, consider the number difference $\Delta
N=\sum_{k\in\Z}kP^\Delta_k$ defined in section 2. All the eigen spaces
$\hi_k:=P_k^\Delta(\hi\otimes\hi)$ are infinite dimensional and
the vectors $\{ |k+n,n\rangle\}_{n\geq\max\{0,-k\}}$ constitute an
orthonormal basis of $\hi_k$. One may thus define a unitary
operator $D$ on $\hi\otimes\hi$ so that, for each $k\in\Z$,
$D(\hi_k)=\hi_{k-1}$. To exhibit such an operator we rename the
basis vectors using the notation of Ban:
$$
|k,n\rangle\rangle:=\left\{\begin{array}{ll}
|n+k,n\rangle, & k\geq 0 \\
|n,n-k\rangle, & k< 0.\end{array}\right.
$$
Then, for any $k\in\Z$, the spectral projection $P_k^\Delta$ can be expressed as
$P_k^\Delta=\sum_{n\in\N}\ketbrab{k,n}{k,n}$, and one may choose, for instance,
$$
D=\sum_{k\in\Z}\sum_{n\in\N}\ketbrab{k-1,n}{k,n}.
$$
This is Ban's proposal for a phase operator.
Writing
\begin{equation}\label{BanD}
D=\int_0^{2\pi}e^{i\theta}\ \ud B(\theta),
\end{equation}
the spectral measure $B$ of $D$ has the form
$$
B(X) =\sum_{k,l\in\Z}\sum_{n\in\N}\frac{1}{2\pi}\int_X
e^{i(k-l)\theta}\ud\theta\ \ketbrab{k,n}{l,n}.
$$
Clearly, $B$ is not phase shift covariant so that it is not a phase observable
in the sense of \cite{Holevo} or \cite{OQP}. However,
the spectral measure $B$ fullfills the covariance condition
$$
V_{\Delta}(\beta)B(X)V_{\Delta}(\beta)^*=B(X\dotplus\beta)
$$
for all $\beta\in\R, X\in\bo$. This differs from the covariance
condition (\ref{kovarianssi}) by the factor 2. Thus $B$ is neither a phase
difference observable in the sense of Definition~\ref{def1}.
The difference by the factor 2 in the
covariance conditions satisfied by $B$ and $\Ecan$ is also reflected
in the commutation
properties of $D$ and $C^{(1)}_{\Ecan}$
with $\Delta N$. Indeed, for all $k\in\Z$ and $n\in\N$,
\begin{equation}\label{kommu1}
[D,\Delta N] |k,n\rangle\rangle = D |k,n\rangle\rangle
\end{equation}
whereas
\begin{equation}\label{kommu2}
[C^{(1)}_{\Ecan},\Delta N] |k,n\rangle\rangle = 2C^{(1)}_{\Ecan}
|k,n\rangle\rangle.
\end{equation}
Notice also that the first cyclic moment $C^{(1)}_{E_{{\rm
can}}}$ of the canonical phase observable $E_{{\rm can}}$
satisfies
\begin{equation*}
[C^{(1)}_{E_{{\rm can}}},N] |n\rangle=C^{(1)}_{E_{{\rm
can}}}|n\rangle
\end{equation*}
for all $n\in\N$. The factor 2 in the covariance condition
(\ref{kovarianssi}) and the commutation relation (\ref{kommu2}) is
natural for a phase difference observable. It is also worth to note
that condition (\ref{kovarianssi}) has a projection valued
solution \cite{italy}. The corresponding unitary operator is
$$
\sum_{k\in\Z}\sum_{n\in\N}\ketbrab{k-2,n}{k,n}.
$$
Compared to (\ref{BanD}), here is again 2 instead of 1.

Although spectral measure $B$ is neither a phase observable nor a
phase difference observable, it has the following relation to
canonical phase observable. When the second mode is in the vacuum
state and the first mode is in an arbitrary state $T$, then
$$
{\rm tr}\left(T\otimes|0\rangle\langle0|\,B(X)\right)
={\rm tr}\left(TE_{\rm can}(X)\right),\;\;\;\;X\in\bo.
$$

\section{Discussion}

The first phase difference operators studied in the literature were suggested by
Sussking and Glogover \cite{SuGl} (see also \cite{CaNi,Vol}).
Their operators were the so-called cosine and sine phase difference operators which
can be represented as $C_{12}=\int_{0}^{2\pi}\cos\theta\,\ud E_{\rm can}^{\rm diff}(\theta)$ and
$S_{12}=\int_{0}^{2\pi}\sin\theta\,\ud E_{\rm can}^{\rm diff}(\theta)$, respectively.
The operators $C_{12}$ and $S_{12}$ do not commute,
and their spectra are the inteval $[-1,1]$, including
a countable dense set of eigenvalues \cite{SuGl,CaNi,Vol}.

L\'evy-Leblond \cite{Le} defined the relative exponential phase operator
$\int_0^{2\pi}e^{i\theta}\ud E_{\rm can}(\theta)\otimes
\int_0^{2\pi}e^{-i\theta}\ud E_{\rm can}(\theta)=
\int_0^{2\pi}e^{i\theta}\ud E_{\rm can}^{\rm diff}(\theta)$
by analogy with the classical expression $e^{i(\theta_1-\theta_2)}
=e^{i\theta_1}e^{-i\theta_2}$.
The operator $\int_0^{2\pi}e^{i\theta}\ud E_{\rm can}^{\rm diff}(\theta)$
is not unitary
but it is associated with the polar decompositon of
$a\otimes a^*$ in the following way: using $|a\otimes a^*|=\sqrt{N\otimes(N+I)}$,
$$
a\otimes a^*=\int_0^{2\pi}e^{i\theta}\ud E_{\rm can}^{\rm diff}(\theta)\,\sqrt{N\otimes(N+I)}.
$$
We can add an extra operator $T:=\sum_{n=0}^\infty\kb n 0 \otimes
\kb 0 n$ to $\int_0^{2\pi}e^{i\theta}\ud E_{\rm can}^{\rm
diff}(\theta)$ and it still satisfies the polar decomposition
relation of $a\otimes a^*$. When doing this we get a unitary
operator $\mathcal E_{12}:=\int_0^{2\pi}e^{i\theta}\ud E_{\rm
can}^{\rm diff}(\theta)+T$ and, thus, a self-adjoint operator
$\Phi_{12}$ such that $\mathcal E_{12}=e^{i\Phi_{12}}$. Obviously,
the operator $\Phi_{12}$ is not the first moment operator of a covariant
phase difference observable. Luis and S\'anchez-Soto have shown
\cite{LuSa1} that the point spectrum of $\Phi_{12}$ consists of
eigenvalues $\left\{2\pi
r/(n+1)\,\big|\,n\in\N,\;r=0,1,...,n\right\}\subset[0,2\pi)$, the closure 
of this set being $[0,2\pi]$.
When the second mode is in a large amplitude coherent state $\ket
z$, the spectral measure of $\Phi_{12}$ gives essentially the same
results as $E_{\rm can}$ (or the difference of $E_{\rm can}$ 
and $F_{\arg z}$) \cite{LuSa1, DaLa}.

Finally, we note that  in an eight-port homodyne detection  the measurement data
is always discrete. Only in the 
limit of large intesity  of the known
fixed-phase reference oscillator  the data 
becomes (essentially) "continuous"  giving rise to the phase observable $E_{\ket 0}$. 
Thus, strictly speaking 
 eight-port homodyne detection cannot be described
as a measurement of the phase difference observable in {\it two} arbitrary signal fields.
However, using two eight-port homodyne detectors with the same large amplitude fixed-phase reference field
 one can measure the difference 
of the two phase observables $E_{\ket 0}$ and $E_{\ket 0}$ \cite{ToMa}.


\end{document}